\documentclass[aps,amsmath,amssymb,reprint]{revtex4-2}
\usepackage{graphicx}%
\usepackage{booktabs}%
\usepackage{overpic}
\usepackage{epsfig}
\usepackage{hyperref} \hypersetup{colorlinks,linkcolor=blue,citecolor=blue}

\newcommand{\pip}{\pi^+}
\newcommand{\pim}{\pi^-}
\newcommand{\acp}{A_{C\!P}}
\newcommand{\dc}{d'_{c}}
\newcommand{\pipi}{\pi^{+}\pi^{-}}
\newcommand{\psip}{\psi(3686)}
\newcommand{\jpsi}{J/\psi}
\newcommand{\EE}{e^+e^-}
\newcommand{\MM}{\mu^+\mu^-}
\newcommand{\LL}{l^+l^-}
\usepackage{lineno}
\newcommand{\BESIIIorcid}[1]{\href{https://orcid.org/#1}{\hspace*{0.1em}\raisebox{-0.45ex}{\includegraphics[width=1em]{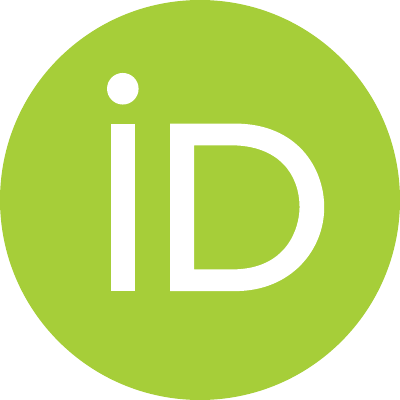}}}}

\begin{document}

\title[Article Title]{Precise Measurement of Chromoelectric Dipole Moment of the Charm Quark}
\author{BESIII Collaboration}
\thanks{Full author list given at the end of the Letter}

\begin{abstract}
The combined symmetry of charge conjugation and parity ($C\!P$) is tested in the hadronic transition $\psip\to\pipi\jpsi$, using a dataset of $(2712.4\pm14.3)\times10^6$ $\psip$ events collected by the BESIII detector at the BEPCII collider. 
The resulting asymmetry observable is $\acp = (0.6\pm1.8\pm0.1)\times10^{-4}$, which is determined by combining the two channels $\jpsi\to\EE$ and $\jpsi\to\MM$. 
Additionally, by considering the relationship between the chromoelectric dipole moment (CEDM) and the $\acp$ observable derived from the quantum chromodynamics multipole expansion theory based on the Chen-Kuang model and the Cornell potential model, we obtain the results of charm quark's CEDM with $\dc = (2.6\pm7.8\pm0.4\pm0.6)\times10^{-16}$ $e$ cm, and $\dc = (3.5\pm10.5\pm0.6\pm0.5)\times10^{-16}$ $e$ cm, respectively. The uncertainties are statistical, systematic, and theoretical, respectively.
These results correspond to an upper limit of $|\dc|<2.1\times10^{-15}$ $e$ cm at 90\% confidence level, representing an order of magnitude improvement in sensitivity compared to the previous direct bound using the same decay process.
\end{abstract}

\maketitle
The existence of a nonzero permanent electric dipole moment (EDM), $d$, via the energy shift $U=-\vec{d}\cdot\vec{E}$, implies a violation of time-reversal symmetry~\cite{Purcell:1950zz, Pospelov:2005pr}.
Here, $\vec{d}=2d\vec{s}/\hbar$ for a nondegenerate particle, $\vec{E}$ is the effective electric field, $\vec{s}$ is the spin, and $\hbar$ is the reduced Planck constant.
Investigations of EDMs enrich our understanding of natural symmetries and open up new avenues for research into the behavior of particles at a fundamental level.
According to the $C\!PT$ theorem for quantum field theories, any violation of the time-reversal indicates a corresponding violation of $C\!P$ symmetry~\cite{Luders:1954zz, Pauli:1956zz}. 
Although $C\!P$ violation has been well established in strange, beauty, and charm meson decays~\cite{Christenson:1964fg, Aubert:2001nu, Abe:2001xe, Aaij:2019kcg}, all measurements to date are consistent with the Standard Model (SM) and are insufficient to fully account for the observed matter-antimatter asymmetry in the universe.
This motivates the search for new sources of $C\!P$ violation, particularly through precise measurements of the EDM of elementary particles and nucleons~\cite{Pospelov:2005pr, Ramsey:1982pq, Ema:2022yra}.

In experiments, the measured neutron EDM is found to be remarkably small~\cite{Abel:2020pzs}. The resulting stringent upper limit for the $C\!P$ violation in QCD processes is known as the strong $C\!P$ problem. It has given rise to searches for a Goldstone boson, the axion~\cite{Wilczek:1977pj, Peccei:1977hh}, which also serves as an attractive candidate for dark matter~\cite{deSwart:2017heh}.
The neutron EDM could originate from the contributions of quark EDMs and chromoelectric dipole moments (CEDMs)~\cite{Gunion:1990iv, DERUJULA1990640}, including those from the second-generation strange and charm quarks~\cite{Sala:2013osa}.
More precise determinations of the EDMs and CEDMs of quarks are essential to understand the neutron EDM and to look for new physics beyond the SM~\cite{Ema:2022pmo}.

\begin{figure}[!htbp]
    \centering
    \includegraphics[width=0.48\textwidth]{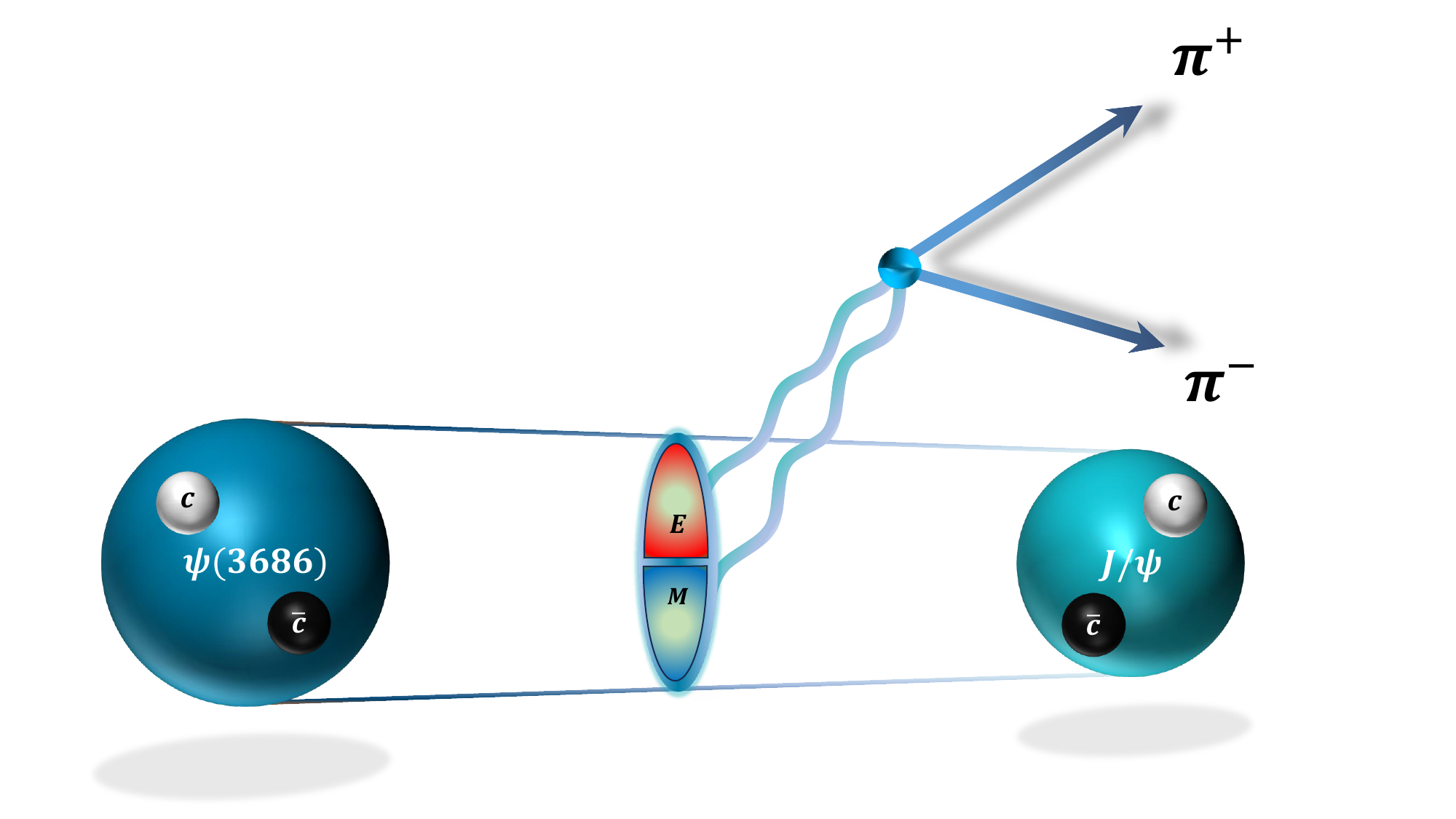}
    \caption{Diagram of $\psip\to\pipi\jpsi$ decay process. The $E$ indicates the color-electric dipole interaction, whereas the $M$ indicates the magnetic dipole interaction that implies the contribution from CEDM.}
    \label{fig:fermion}
\end{figure}

The CEDM arises from the anomalous coupling between quarks and gluons, which is described by the effective Lagrangian~\cite{Kuang:2012wp}
\begin{equation}
    \mathcal{L}_{\rm CEDM} = -i/2 \dc \bar{\psi}_c \sigma^{\mu\nu} \gamma_5 \lambda_a/2 \psi_c G^a_{\mu\nu},
\end{equation}
where $\dc$ is the strength of the CEDM, $\sigma^{\mu\nu}=i/2[\gamma^\mu, \gamma^\nu]$, $\lambda_a$ are the Gell-Mann matrices of the color SU(3)$_c$ group, and $G_{\mu\nu}^a$ is the gluon field strength tensor. 
In the charmonium system ($c\bar{c}$), hadronic transitions such as $\psip\to\pipi\jpsi$ are characterized by the electric form factors corresponding to $S$- and $D$-wave $C\!P$-even states~\cite{Kuang:1989ub,Kuang:2002hz}.
A nonzero CEDM induces a magnetic form factor through mixing with the $C\!P$-odd $P$-wave contribution~\cite{Kuang:2012wp}.
Consequently, the $\psip\to\pipi\jpsi$ transition (see Fig.~\ref{fig:fermion}) provides a unique probe of $C\!P$ symmetry via the mixing of $C\!P$-even and $C\!P$-odd charmonium states in the quantum chromodynamics multipole expansion (QCDME)~\cite{Kuang:2012wp,Yan:1980uh,Kuang:2006me}.

Within the QCDME framework, $\dc$ is proportional to the $C\!P$-odd observable $\acp$, which can be parameterized as~\cite{Kuang:2012wp}
\begin{eqnarray}
    \acp &=& (9.33\pm2.34) \times 10^{-3} \dc  m_{c}\ [\text{CK model}], \label{eq:CK}\\ 
    \acp &=& (6.93\pm0.97) \times 10^{-3} \dc  m_{c}\ [\text{Cornell model}], \label{eq:CN}
\end{eqnarray}
where Eq.~\eqref{eq:CK} is based on the Chen-Kuang (CK) model for the heavy quark potential~\cite{Chen:1992fq}, and Eq.~(\ref{eq:CN}) corresponds to the simpler Cornell potential model~\cite{Eichten:1978tg}. Here, $m_{c}$ denotes the mass of the charm quark~\cite{ParticleDataGroup:2024cfk}.

 A nonzero $\acp$ value in this hadronic transition process would signal physics beyond the SM and could constitute the first observation of a nonvanishing CEDM.
An indirect upper limit on the charm CEDM, $|\dc|<1\times10^{-22}$ $e$ cm, has been inferred from neutron EDM measurements~\cite{Sala:2013osa}.
A directly obtained upper limit of $\dc<3\times10^{-14}$ $e$ cm was reported in a phenomenological study of hadronic transition $\psip\to\pipi\jpsi$ using data from the BESII experiment~\cite{Kuang:2012wp, BES:2006eer}. Among quarks, the top quark remains the other one for which a direct experimental bound on its CEDM has been established by the CMS Collaboration~\cite{CMS:2022voq}. To date, there has been no precise experimental measurement of the charm quark's CEDM.
The world's largest $\psip$ data sample collected by the BESIII detector~\cite{BESIII:2024lks}, together with the large branching fraction $\text{BR}[\psip\to\pipi\jpsi]=(34.64\pm0.34)\%$~\cite{ParticleDataGroup:2024cfk}, enables a direct measurement of the CEDM and a sensitive test of strong $C\!P$ symmetry.

In this Letter, we conduct a comprehensive investigation of $C\!P$ violation
in charm quark-antiquark bound states via the hadronic transition $\psip\to\pipi\jpsi$,
using a data sample of $(2712.4\pm14.3)\times10^6$ $\psip$ events collected by the BESIII detector at the BEPCII collider.
The leptonic decay modes $\jpsi\to \LL$, where $l=e,\,\mu$, are used. These constitute the largest $\jpsi$ decay channels, with branching fractions of approximately 6\% each~\cite{ParticleDataGroup:2024cfk}.
The design and performance of the BESIII detector are described in Refs.~\cite{BESIII:2009fln, BES:2001vqx, Yu:2016cof}.
 
A signal Monte Carlo (MC) simulation of the decay process $\psip\to\pipi\jpsi$, containing 42.8 million events for each $\jpsi$ decay channel $\jpsi\to l^+l^-$, is generated to determine the detection efficiency and optimize event selection criteria. The simulation is carried out using a {\sc geant4}-based~\cite{GEANT4:2002zbu} MC package, which includes the geometric description of the BESIII detector and the detector response. The simulation models the beam energy spread and initial state radiation (ISR) in $e^+e^-$ annihilation with the {\sc kkmc} generator~\cite{Jadach:2000eu}.
Particularly, the decay process $\psip\to\pipi\jpsi$ is modeled using a $\pi\pi$ transition amplitude approach, incorporating chiral perturbation theory and final state interactions~\cite{Mannel:1995jt,Oller:1997ti,Guo:2004dt,Liu:2012dv}.  
To improve the consistency between data and MC simulation, a tracking efficiency correction for the two pions is applied to the signal MC samples.
The efficiencies are determined from two-dimensional distributions of transverse momentum versus polar angle using the sample of $\psip\to\pipi\jpsi$ with $\jpsi\to\EE$.
An additional inclusive MC sample including the production of the $\psi(3686)$ resonance, the ISR production of the $J/\psi$, and the continuum processes incorporated in {\sc kkmc} is used to estimate the possible backgrounds.
All particle decays are modeled with {\sc
evtgen}~\cite{Lange:2001uf,Ping:2008zz} using branching fractions 
either taken from the
Particle Data Group~\cite{ParticleDataGroup:2024cfk}, when available,
or otherwise estimated with {\sc lundcharm}~\cite{Chen:2000tv,Yang:2014vra}.
Final state radiation
from charged final state particles is incorporated using the {\sc
photos} package~\cite{Barberio:1990ms}.

Charged tracks detected in the multilayer drift chamber (MDC) are required to be within a polar angle ($\theta$) range of $|\cos\theta|<0.93$, where $\theta$ is defined with respect to the $z$ axis, which is the symmetry axis of the MDC. The distance of closest approach of the charged tracks to the $\EE$ interaction point (IP) must be less than 10~cm along the $z$ axis, and less than 1~cm in the transverse plane. 
Pions and leptons (electrons and muons) are distinguished based on momentum ($p$) criteria derived from the MDC: two high-momentum tracks ($p>1.0$ GeV/$c$) are identified as leptons, and two low-momentum tracks ($p<1.0$ GeV/$c$) are identified as pions.
To further separate electrons and muons, leptons or antileptons with deposited energies in the electromagnetic calorimeter (EMC) greater than 1.0 GeV are categorized as electrons or positrons, while those with energies less than 0.45 GeV are classified as muons. Given that the final state for our signal process corresponds to either $\pipi\EE$ or $\pipi\MM$, we require four charged tracks per candidate event having zero net charge comprising one $\pip$, one $\pim$, one $e^+$ ($\mu^+$), and one $e^-$ ($\mu^-$) for $\jpsi \to \EE$ or $\jpsi \to \MM$.
To ensure that all final states originate from the IP, a primary vertex fit is performed on the four charged tracks corresponding to $\pipi\LL$. 
Following the above requirements, a four-constraint (4C) kinematic fit is applied to combinations of the final states, enforcing energy-momentum conservation. 
The $\chi^{2}_{\rm 4C}$ of the kinematic fit is required to be less than 200 to improve mass resolution and suppress potential background contributions. The selection criterion for $\chi^2_{\rm 4C}$ is validated using a figure of merit (FOM) defined as $S/\sqrt{S+B}$, where $S$ represents the number of signal events estimated from the MC simulation, and $B$ denotes the background contribution. The quantity $S+B$ is derived from data events.
An optimized mass window on the $\LL$ invariant mass distribution, $|m_{\LL}-M_{\jpsi}|<0.01$ GeV/$c^2$, is imposed to select $\jpsi$ candidates based on the FOM method. The $m_{\LL}$ is the invariant mass of leptons, and $M_{\jpsi}$ is the nominal mass of $\jpsi$~\cite{ParticleDataGroup:2024cfk}.

 To study background contributions from continuum processes, we analyze $\EE\to\pipi\jpsi$ events collected at $\sqrt{s}=3.65$ GeV with an integrated luminosity of 401 pb$^{-1}$~\cite{BESIII:2024lks}. 
To suppress such backgrounds, we require $|\cos\theta_{\pi l}|>0.95$ and $m_{\pipi}<0.32$ GeV/$c^2$, where $\cos\theta_{\pi l}$ represents the opening angle between $\pip$ and $l^-$, and $m_{\pipi}$ denotes the invariant mass of the $\pipi$ system. 
The remaining continuum events in the signal region are normalized 
according to the CM energy and luminosity. The normalization factor is 
     $\frac{\mathcal{L}_{\psip}}{\mathcal{L}_{3650}}\times\frac{s_{3650}}{s_{\psip}}$,
where $\mathcal{L}_{\psip}$ ($\mathcal{L}_{3650}$) is the integrated luminosity of the $\psip$ (continuum) data sample at the CM energy of $\sqrt{s_{\psip}}$ ($\sqrt{s_{3650}}$).
After applying the same event selection criteria to the inclusive MC sample~\cite{Zhou:2020ksj}, a comprehensive study of potential background contributions indicates that all remaining backgrounds in the signal region are negligible.
The final signal yields are $(2.160\pm0.001)\times10^{7}$ and $(2.937\pm0.001)\times10^{7}$,
with estimated continuum background fractions of 0.04\% and 0.01\% for $\jpsi\to\EE$ and $\jpsi\to\MM$ channels, respectively.

\begin{figure*}[!htbp]
			\centering
			\includegraphics[width=1.0\textwidth]{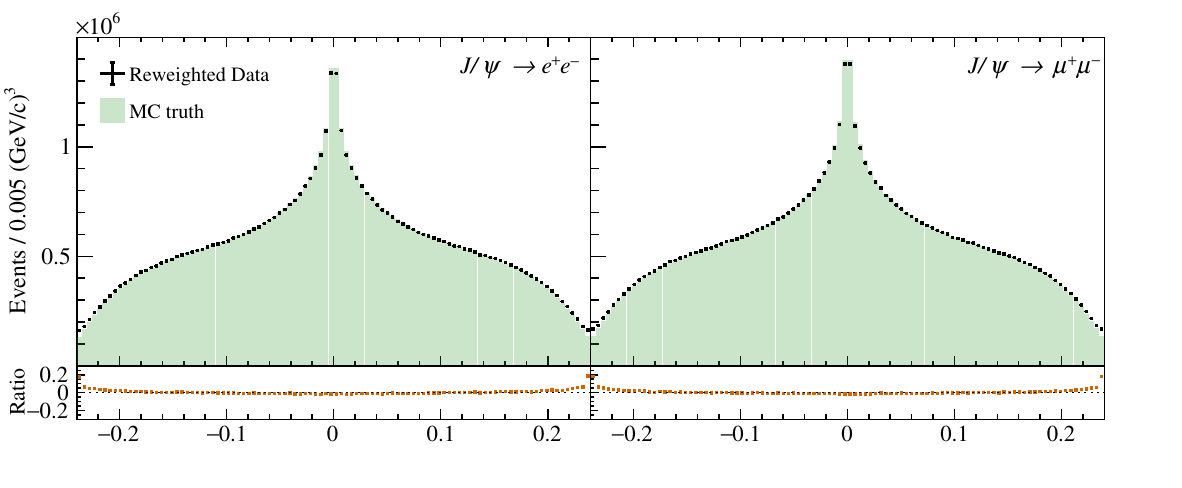} \put(-245,2){$q_{3}$ $(\mathrm{GeV}/c)^3$}
			\caption{Distributions of $C\!P$-odd observable $q_{3}$. The black dots with error bars are the effective data events reweighted by efficiency curves; the green bar chart region denotes the MC truth. The left and right figures correspond to $\jpsi\to\EE$ and $\jpsi\to\MM$ decay channels, respectively. The ratio in the bottom subplot is defined as $\rm 1-{N_{\rm truth}}/{N_{\rm events}}$. The events in the range of $|q_{3}|<0.24~(\mathrm{GeV}/c)^3$ are kept.}
			\label{fig:q3}
\end{figure*}

To characterize the $C\!P$ asymmetry, a $C\!P$ observable $q_{3}$ is defined as~\cite{Kuang:2012wp}
\begin{equation}
    q_{3} \equiv \vec{p}_{e^+} \cdot (\vec{p}_{\pip} - \vec{p}_{\pim}) \vec{p}_{e^+} \cdot \frac{\vec{p}_{\pip}\times \vec{p}_{\pim}}{|\vec{p}_{\pip}\times \vec{p}_{\pim}|},
    \label{eq:q3}
\end{equation}
where $\vec{p}_{e^+},\ \vec{p}_{\pip}$ and $\vec{p}_{\pim}$ represent the momentum vectors of beam $e^+$ and the final states $\pi^+$ and $\pi^-$ in the $\EE$ center-of-mass frame. 
The observable $q_{3}$ is $C\!P$-odd, satisfying the $C\!P$ transformation condition $\boldsymbol{\mathrm{ C\!P}} (q_{3}) = -q_{3}$.
The effective $C\!P$ asymmetry observable is defined as
\begin{eqnarray}
    \acp 
    = \frac{N_{\rm events}(q_{3}>0) - N_{\rm events}(q_{3}<0)}{N_{\rm events}(q_{3}>0) + N_{\rm events}(q_{3}<0)}.
    \label{eq:acp}
\end{eqnarray}
The $\acp$ value directly quantifies the $C\!P$-odd asymmetry in the distribution of the observable $q_3$.
The signal yields for each bin $N_{\rm events}^i$ in the $q_{3}$ distribution, with the bins ranging over $i=[1, 100]$, which account for the effects of detector and reconstruction biases, are calculated from the efficiency obtained from the simulations.
This can be expressed as $N_{\rm events}^i = {N^{i}_{\rm obs}}/{\epsilon_i}$.
Here $N^{i}_{\rm obs}$ is the number of signal events observed for each bin,
the efficiency for each bin $i$, denoted as $\epsilon_i = {N_{\rm rec}^i}/{N_{\rm truth}^i}$, is defined as the ratio of reconstructed to generated MC events.
Figure~\ref{fig:q3} illustrates the $q_{3}$ distributions of the generated MC sample and the reweighted data for both $\jpsi$ decay channels. 
To extract the signal yields in the two regions of $q_{3}<0$ and $q_{3}>0$ as shown in Fig.~\ref{fig:q3}, a counting method is used, which allows the determination of the asymmetry $\acp$ using Eq.~(\ref{eq:acp}). The numerical results for the asymmetries $\acp$ are then summarized in Table~\ref{tab:result} for decays $\jpsi\to \LL$. Additionally, an average value $\langle\acp\rangle$ is calculated by combining the two channels, while the correlations are treated according to Ref.~\cite{Schmelling:1994pz}.

\begin{table}[htbp]
    \centering
    \caption{The measured $C\!P$ observables and CEDM measurement. The second line represents the average value of $\acp$. The CEDM $\dc$ results from two assumptions of potential functions~\cite{Chen:1992fq,Eichten:1978tg}. The first uncertainties are statistical, the second systematic, and the third from the theoretical model~\cite{Kuang:2012wp}.
    The upper limit of $\dc$ is determined at a 90\% confidence level. }
    \begin{tabular}{lcc}
    \toprule
        Parameter & $\jpsi\to\EE$ & $\jpsi\to\MM$ \\ \midrule
        $\acp$ ($\times10^{-4}$) & $3.3 \pm 2.9 \pm 0.4$   &   $-1.2 \pm 2.4 \pm 0.1$ \\
        $\langle\acp\rangle$ ($\times10^{-4}$) & \multicolumn{2}{c}{$0.6\pm1.8\pm0.1$} \\
        $\dc$ [CK] ($e$ cm) & \multicolumn{2}{c}{$(2.6\pm7.8\pm0.4\pm0.6)\times10^{-16}$} \\
        $\dc$ [Cornell] ($e$ cm) & \multicolumn{2}{c}{$(3.5\pm10.5\pm0.6\pm0.5)\times10^{-16}$} \\ 
        $|\dc|$ ($e$ cm)&  \multicolumn{2}{c}{$<2.1\times10^{-15}$} \\
        \botrule
    \end{tabular}
    \label{tab:result}
\end{table}

The sources of systematic uncertainties associated with the $\acp$ measurements arise from discrepancies between data and MC simulation, including tracking, 4C kinematic fit, the $\jpsi$ mass window, and background contributions. 
The systematic uncertainties related to $\EE$ and $\MM$ tracking are investigated using control samples from the radiative Bhabha process $\jpsi \to \gamma \EE$~\cite{Chai:2025xni} and $\jpsi \to \pi^+ \pi^- \pi^0$~\cite{Liu:2024uno}. The muon tracking efficiency is taken from that of pions due to their similar mass characteristics. The difference between results corrected for efficiencies and those without corrections is considered a systematic uncertainty. 
To evaluate the contribution of the 4C kinematic fit, we modify the helix track parameters for all final state particles following the methodology outlined in Ref.~\cite{BESIII:2012mpj}. In our nominal analysis, we apply this correction; however, to evaluate its impact, we also analyze results obtained without this correction and take their difference as a source of systematic uncertainty. 
Because of the bremsstrahlung of electrons, the kinematic systematic uncertainty of electrons is relatively large ($0.3\times10^{-4}$), while the error of muons in this regard is negligible.
The systematic uncertainties arising from the $\jpsi$ selection have been investigated by varying the selection regions~\cite{Barlow:2002yb}. The $\jpsi$ mass window is adjusted within a range of 7 to 14 MeV/c$^2$, while the $q_{3}$ requirement is varied from 0.21 to 0.28$~(\mathrm{GeV}/c)^3$. It was observed that the $\acp$ values remain unaffected within a threshold of 2$\sigma$. Consequently, the systematic uncertainties associated with the $\jpsi$ mass window and the $q_{3}$ requirement are negligible.
To examine effects originating from backgrounds, we focus on continuum processes in the signal region; thus, the differences in $\acp$ measurements with and without background are taken as systematic uncertainties. 
A detailed analysis of systematic uncertainties and the list of these systematic uncertainties are presented in the End Matter.
Assuming independence among all sources, the total systematic uncertainties are derived, by summing the individual contributions in quadrature, to be $0.4\times10^{-4}$ and $0.1\times10^{-4}$ for $\jpsi\to\EE$ and $\jpsi\to\MM$, respectively. 

\begin{figure}[!htbp]
    \centering
    \includegraphics[width=0.5\textwidth]{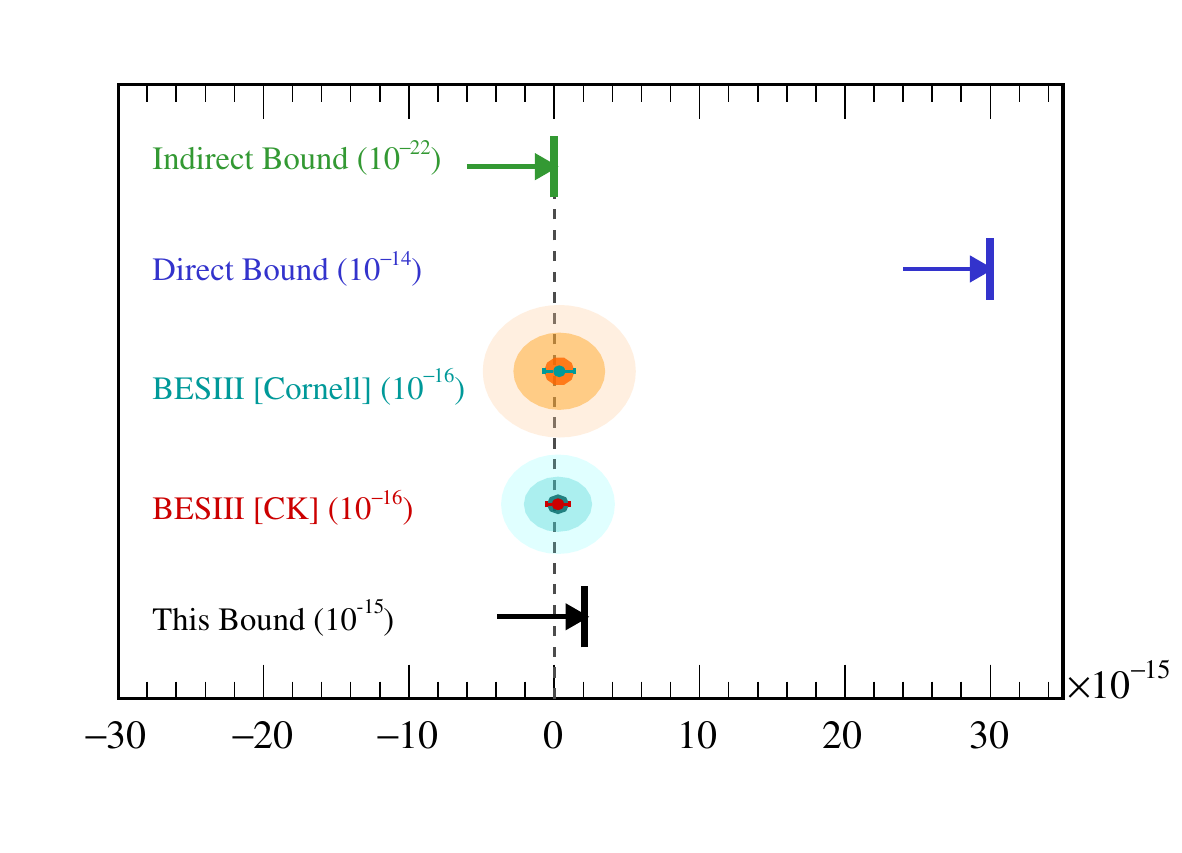}
    \put(-145,5){$\dc$ ($e$ cm)}
    \caption{The measured charm CEDMs and (phenomenological) upper limits. The cyan and red dots, along with their associated error bars, represent the BESIII results under the assumptions of the CK model~\cite{Chen:1992fq} and the Cornell model~\cite{Eichten:1978tg}, respectively. 
    The inner, middle, and outer contours represent 68.2\%, 95.4\%, and 99.7\% confidence levels of the $\dc$, respectively.
    The black arrow limitation implies the upper limit of $\dc$ in this work at a 90\% confidence level.
    The green and blue arrow limitations show the indirect constraint inferred from a neutron EDM measurement~\cite{Sala:2013osa}, and the direct~\cite{Kuang:2012wp} phenomenological bound, respectively. 
    }
    \label{fig:results}
\end{figure}

In conclusion, using a data sample of $(2712.4\pm14.3)\times10^6$ $\psip$ events collected with the BESIII detector, we perform a comprehensive investigation for $C\!P$ violation in the hadronic transition $\psip\to\pipi\jpsi$ with subsequent decays $\jpsi\to \LL$ ($\ell = e, \mu$). The numerical results are summarized in Table~\ref{tab:result}.
A $C\!P$-odd observable, constructed from independent momentum directions of the accelerator beam positrons and the pion final state particles, is used to probe $C\!P$ asymmetry. 
The measured $\acp$ values are consistent with $C\!P$ symmetry, with statistical and systematic uncertainties at the $10^{-4}$ and $10^{-5}$ levels, respectively. The SM continues to provide a good description of particle physics phenomena at this energy scale. 
The average $C\!P$ asymmetry, obtained by combining the results from $\jpsi\to\EE$ and $\jpsi\to\MM$ channels, is determined to be ($0.6\pm1.8\pm0.1)\times10^{-4}$.
Using the relations in Eqs.~(\ref{eq:CK}) and (\ref{eq:CN}), the charm CEDM $\dc$ is extracted from the measured $\acp$ value. 
The charm CEDMs are measured to be 
$\dc = (2.6\pm7.8\pm0.4\pm0.6)\times10^{-16}$ $e$ cm under the assumption of the CK model~\cite{Chen:1992fq}, 
and $\dc = (3.5\pm10.5\pm0.6\pm0.5)\times10^{-16}$ $e$ cm under the assumption of the Cornell model~\cite{Eichten:1978tg}. The uncertainties are statistical, systematic, and theoretical, respectively.
These values are consistent with zero and can be accommodated within the SM, further strengthening confidence in our current understanding of particle physics processes involving charm quarks. 

A comparison between our results and previous phenomenological estimates is illustrated in Fig.~\ref{fig:results}. 
The upper limit on $\dc$ at the 90\% confidence level is approximately one order of magnitude more stringent than the previous direct phenomenological limit based on the BESII dataset~\cite{Kuang:2012wp}. 
This represents the most precise measurement of the charm quark CEDM to date, enabled by the large BESIII data sample and clean experimental environment.

The improved constraints on the charm CEDM provide valuable insight into possible $C\!P$-violating interactions beyond the SM within the QCD framework. This reduction in theoretical uncertainty strengthens the role of the CEDM as a key probe of new physics, thereby enhancing our understanding of the origins of the neutron EDM and its connection to quark-level $C\!P$ violation~\cite{Gunion:1990iv}.
Although the measured charm CEDM does not yet reach the indirect upper limit inferred from neutron EDM measurements~\cite{Sala:2013osa}, the significant improvement opens new opportunities for further investigation and analysis, 
such as examining $P$-wave contributions in other charmonium hadronic transitions~\cite{Ma:2003yf}.
The methodology demonstrated here can be extended to investigate $C\!P$ violation in other heavy quarkonium systems, in particular the beauty sector via decays such as $\Upsilon(2S) \to \pi^+\pi^- \Upsilon(1S)$.

\textit{Acknowledgments}—The BESIII Collaboration thanks the staff of BEPCII~\cite{BEPC:2025xni} and the IHEP computing center for their strong support. This work is supported in part by 
National Key R\&D Program of China under Contracts No. 2023YFA1606000, No. 2023YFA1606704; 
the Fundamental Research Funds for the Central Universities No.
lzujbky-2025-ytA05, No. lzujbky-2025-it06, No. lzujbky-2024-jdzx06;
National Natural Science Foundation of China (NSFC) under Contracts No. 12247101, No. 11635010, No. 11935015, No. 11935016, No. 11935018, No. 12025502, No. 12035009, No. 12035013, No. 12061131003, No. 12192260, No. 12192261, No. 12192262, No. 12192263, No. 12192264, No. 12192265, No. 12221005, No. 12225509, No. 12235017, No. 12361141819; 
the Natural Science Foundation of Gansu Province No. 22JR5RA389, No. 25JRRA799;
the "111 Center" under Grant No. B20063;
the Chinese Academy of Sciences (CAS) Large-Scale Scientific Facility Program; CAS under Contract No. YSBR-101; 100 Talents Program of CAS; The Institute of Nuclear and Particle Physics (INPAC) and Shanghai Key Laboratory for Particle Physics and Cosmology; German Research Foundation DFG under Contract No. FOR5327; Istituto Nazionale di Fisica Nucleare, Italy; Knut and Alice Wallenberg Foundation under Contracts No. 2021.0174, No. 2021.0299; Ministry of Development of Turkey under Contract No. DPT2006K-120470; National Research Foundation of Korea under Contract No. NRF-2022R1A2C1092335; National Science and Technology fund of Mongolia; National Science Research and Innovation Fund (NSRF) via the Program Management Unit for Human Resources \& Institutional Development, Research and Innovation of Thailand under Contract No. B50G670107; Polish National Science Centre under Contract No. 2024/53/B/ST2/00975; Swedish Research Council under Contract No. 2019.04595; U. S. Department of Energy under Contract No. DE-FG02-05ER41374.

\textit{Data availability}—The data that support the findings of this article are not publicly available upon publication
because  it is not technically feasible and/or the cost of preparing, depositing,  and hosting the data would  be prohibitive
within  the terms  of this research project. The data are available from the authors upon reasonable request.

%


\appendix
\section*{End Matter}
This End Matter provides a detailed analysis of the systematic uncertainties for each source, including the kinematic fit, $\jpsi$ selection, $q_3$ range selection, and background estimation.
The contributions from each source and uncertainty, together with the total systematic uncertainty, are summarized in Table~\ref{tab:sysunc}.
\begin{table}[htbp]
    \centering
    \caption{The systematic uncertainties in unit of $10^{-4}$ of the $\acp$ measurements.}
    \begin{tabular}{lcc}
    \toprule
         Source ($\times10^{-4}$)                    & $\jpsi\to\EE$ & $\jpsi\to\MM$ \\ \midrule
        Tracking for $l^{-}$        & 0.1       & $<$ 0.1 \\
        Tracking for $l^{+}$        & $<$ 0.1   & $<$ 0.1 \\
        Tracking for $\pipi$        & $<$ 0.1   & $<$ 0.1 \\ 
        Kinematic fit               & 0.3       & $<$ 0.1 \\
        Background                  & 0.2       & 0.1 \\ \midrule
        Summary                     & 0.4       & 0.1 \\ \botrule
    \end{tabular}
    \label{tab:sysunc}
\end{table}

\textit{Kinematic fit}—The tracks used in the kinematic fit, $\pipi\LL$, are corrected using helix parameters. The systematic uncertainty associated with the kinematic fit is evaluated by comparing results obtained with and without these corrections. The observed difference is small, indicating that the $\acp$ measurements are largely insensitive to the details of the kinematic fit. A larger uncertainty is observed in the electron channel ($0.3\times10^{-4}$) compared to the muon channel ($<0.1\times10^{-4}$), which can be attributed to electron bremsstrahlung effects.

\textit{The $\jpsi$ selection}—The systematic uncertainty arising from the $J/\psi$ selection is studied using the Barlow test, by varying the mass window cut between 7 and 14 MeV/$c^2$.
The significance is evaluated via $\zeta = \Delta a / \sigma_a$, where $\Delta a = |a_{\text{nom}} - a_{\text{syst}}|$ denotes the absolute difference in the parameter $a$ between the nominal and varied selections, with $a_{\text{nom}}$ and $a_{\text{syst}}$ representing their respective central values. The uncorrelated uncertainty is defined as $\sigma_a = \sqrt{|\sigma_{a, \text{nom}}^2 - \sigma_{a, \text{syst}}^2|}$, where $\sigma_{a, \text{nom}}$ and $\sigma_{a, \text{syst}}$ are the corresponding statistical uncertainties, respectively.
All tested variations yield $\zeta < 2.0$, with most below 1.0, indicating no significant deviations from the nominal measurements.

\textit{The $q_3$ range selection}—The selection criterion $q_3$ is based on the slope of the efficiency curve, defined as the ratio of reconstructed to generated events in each bin of the MC sample. Slopes are computed between consecutive $q_3$ points, separated by one data point, and events in regions with large efficiency slopes are excluded.
As shown in Fig.~\ref{fig:q3}, the maximum bias is approximately $20\%$, primarily arising from binning effects at the edges of the distribution where event classification becomes ambiguous. Various binning schemes and range selections were tested, but fully eliminating this bias remains challenging. To assess its impact on the $\acp$ measurement, the Barlow test is performed by varying the range from 0.21 to 0.28$~(\mathrm{GeV}/c)^3$. Nearly all results satisfy $\zeta < 2.0$, with most below 1.5, confirming that no significant deviation is introduced by the $q_3$ selection.

\textit{Background estimation}—The contribution from background events is incorporated into the $\acp$ measurement as follows:
\begin{equation}  
    \acp = \frac{(N_{\rm Data}^+ - N^+_{\rm BKG}) - (N_{\rm Data}^- - N^-_{\rm BKG})}{(N_{\rm Data}^+ - N^+_{\rm BKG}) + (N_{\rm Data}^- - N^-_{\rm BKG})},
\end{equation}  
where $N_{\rm Data}$ and $N_{\rm BKG}$ denote the number of events in data and background samples, respectively. The systematic uncertainty due to background is taken as the difference between results with and without background subtraction. This study focuses on continuum backgrounds, which are found to be higher in the $J/\psi\to e^+e^-$ channel than in the $J/\psi\to\mu^+\mu^-$ channel. Accordingly, the associated systematic uncertainty is larger for the electron channel ($0.2\times10^{-4}$) compared to the muon channel ($0.1\times10^{-4}$).

\begin{widetext}
\begin{center}
\small
M.~Ablikim$^{1}$\BESIIIorcid{0000-0002-3935-619X},
M.~N.~Achasov$^{4,b}$\BESIIIorcid{0000-0002-9400-8622},
P.~Adlarson$^{77}$\BESIIIorcid{0000-0001-6280-3851},
X.~C.~Ai$^{82}$\BESIIIorcid{0000-0003-3856-2415},
R.~Aliberti$^{36}$\BESIIIorcid{0000-0003-3500-4012},
A.~Amoroso$^{76A,76C}$\BESIIIorcid{0000-0002-3095-8610},
Q.~An$^{73,59,\dagger}$,
Y.~Bai$^{58}$\BESIIIorcid{0000-0001-6593-5665},
O.~Bakina$^{37}$\BESIIIorcid{0009-0005-0719-7461},
Y.~Ban$^{47,g}$\BESIIIorcid{0000-0002-1912-0374}
% ...
\\
\vspace{0.2cm}
(BESIII Collaboration)\\
\vspace{0.2cm} {\it
$^{1}$ Institute of High Energy Physics, Beijing 100049, People's Republic of China\\
% ...
$^{4}$ Budker Institute of Nuclear Physics SB RAS (BINP), Novosibirsk 630090, Russia\\
% ...
$^{36}$ Johannes Gutenberg University of Mainz, Johann-Joachim-Becher-Weg 45, D-55099 Mainz, Germany\\
$^{37}$ Joint Institute for Nuclear Research, 141980 Dubna, Moscow region, Russia\\
% ...
$^{47}$ Peking University, Beijing 100871, People's Republic of China\\
% ...
$^{58}$ Southeast University, Nanjing 211100, People's Republic of China\\
$^{59}$ State Key Laboratory of Particle Detection and Electronics, Beijing 100049, Hefei 230026, People's Republic of China\\
% ...
$^{73}$ University of Science and Technology of China, Hefei 230026, People's Republic of China\\
% ...
$^{76}$ University of Turin and INFN, (a)University of Turin, I-10125, Turin, Italy; (b)University of Eastern Piedmont, I-15121, Alessandria, Italy; (c)INFN, I-10125, Turin, Italy\\
$^{77}$ Uppsala University, Box 516, SE-75120 Uppsala, Sweden\\
% ...
$^{82}$ Zhengzhou University, Zhengzhou 450001, People's Republic of China\\

\vspace{0.2cm}
$^{\dagger}$ Deceased.\\
% ...
$^{b}$ Also at the Novosibirsk State University, Novosibirsk, 630090, Russia.\\
% ...
$^{g}$ Also at State Key Laboratory of Nuclear Physics and Technology, Peking University, Beijing 100871, People's Republic of China.\\
% ...
}
\end{center}
\end{widetext}

\end{document}